\documentclass[10pt,twocolumn,english,aps,manuscript,prl]{revtex4}
\usepackage{mathptmx}
\usepackage{helvet}

\usepackage[T1]{fontenc}
\usepackage[latin9]{inputenc}
\usepackage{amssymb}
\usepackage{graphicx}

\makeatletter
\@ifundefined{textcolor}{}
{%
 \definecolor{BLACK}{gray}{0}
 \definecolor{WHITE}{gray}{1}
 \definecolor{RED}{rgb}{1,0,0}
 \definecolor{GREEN}{rgb}{0,1,0}
 \definecolor{BLUE}{rgb}{0,0,1}
 \definecolor{CYAN}{cmyk}{1,0,0,0}
 \definecolor{MAGENTA}{cmyk}{0,1,0,0}
 \definecolor{YELLOW}{cmyk}{0,0,1,0}
}

\PassOptionsToPackage{caption=false}{subfig}

\@ifundefined{showcaptionsetup}{}{%
 \PassOptionsToPackage{caption=false}{subfig}}
\usepackage{subfig}
\makeatother

\usepackage{babel}
\begin{document}

\title{Dramatic changes in electronic structure revealed by fractionally
charged nuclei}

\author{Aron J. Cohen}

\affiliation{Department of Chemistry, Lensfield Rd, University of Cambridge, Cambridge,
CB2 1EW, UK }

\author{Paula Mori-Sánchez}

\email{paula.mori@uam.es}

\affiliation{Departamento de Química, Universidad Autónoma de Madrid, 28049, Madrid,
Spain}
\begin{abstract}
Discontinuous changes in the electronic structure upon infinitesimal
changes to the Hamiltonian are demonstrated. Remarkably, these are
revealed in one and two electron molecular systems if the realm of
the nuclear charge is extended to be fractional. Dramatic changes
in the electron density from full configuration interaction are observed
in real space illustrating key intricacies of electronic structure
including the transfer, hopping and removal of electrons. Physically,
this is due to the particle nature of electrons and manifests itself
theoretically as a diverging linear density response function or an
energy derivative discontinuity that occurs at constant number of
electrons. This is essential to correctly describe real physical processes,
from chemical reactions to electron transport and metal-insulator
transitions. The dramatic errors of DFT are seen in real space as
this physics is missing from currently used approximations and poses
a great challenge for the development of new electronic structure
methods.
\end{abstract}
\maketitle

\section{introduction}

How electrons move upon a change in the external potential, $v(\mathbf{r})$,
is a key question in the understanding of the quantum nature of electrons
in matter, given by the Schrödinger equation
\[
\left(\sum_{i}-\frac{1}{2}\nabla_{i}^{2}+v(\mathbf{r}_{i})+\sum_{i>j}\frac{1}{r_{ij}}\right)\Psi=E\Psi.
\]
A deeper understanding of the basic behaviour of electrons is needed
and is the focus of this work. The change in $v(\mathbf{r})$ in processes
such as stretching bonds, chemical reactions and electron transport
is a great challenge for electronic structure theory. Methods such
as Hartree-Fock (HF) and MP2 work well for many properties such as
equilibrium structures, where the electronic structure is dominated
by a single determinant, however they break down when the basic description
of the wavefunction needs more than one determinant. Currently, the
only way to tackle this challenge is with multi-reference methods
leading to an exact diagonalization of the full Hilbert space with
Full Configuration Interaction (FCI) \cite{booth09054106}, where
the limitation is the exponentially scaling size of the space. Density
Functional Theory (DFT) attacks the same problem in a fundamentally
different manner using the real-space electron density as the fundamental
variable, with all the complexity now hidden in the exchange-correlation
functional $E_{xc}[\rho({\bf r})]$. The same functional has to correctly
describe all systems, i.e. the result of the functional on many systems
is equivalent to many FCI calculations. The simplest example that
this is a non-trivial problem is the incapability to make one functional
that describes the energy of both stretched H$_{2}^{+}$ and stretched
H$_{2}$ \cite{Cohen12289}. From a FCI perspective these two systems
are trivial as they have one and two electrons, however in DFT it
is the use of the same functional that links them (and in fact all
other systems) together that poses a distinct challenge.

\begin{figure}[b]
\includegraphics[scale=0.3]{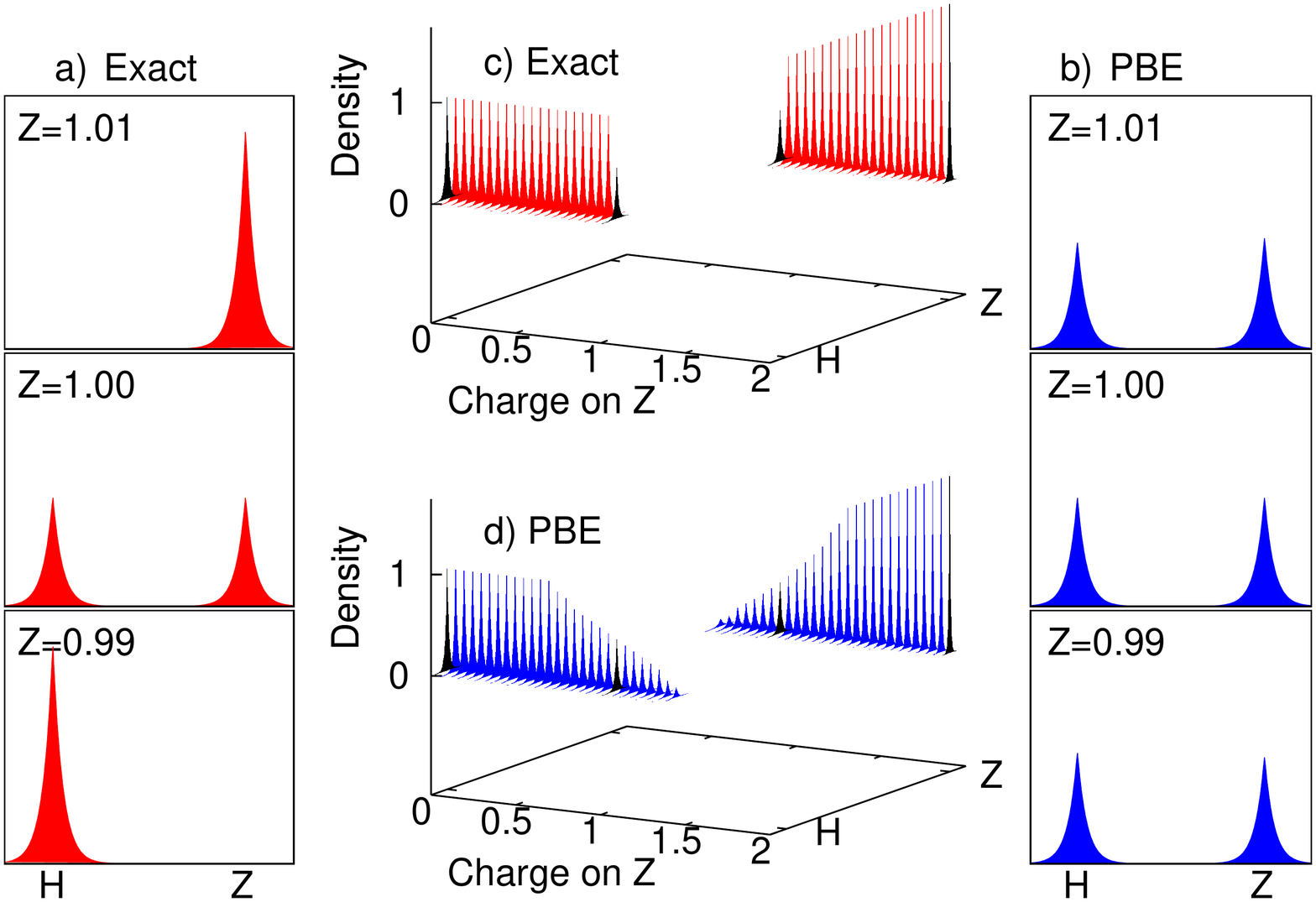}

\caption{HZ$^{(1e)}$ : two protons with one electron, and varying the charge
of one of the protons, $0\le Z\le2$. The density is shown for (a)
Exact and (b) PBE for three charges on the Z atom (0.99,1.00,1.01).
The exact behavior shows very discontinuous behaviour at $Z=1$ that
a functional such as PBE fails to capture due to delocalization error.
In (c) and (d) the same is illustrated but showing the whole range
$0\le Z\le2$, the curves in black correspond to the density for the
integer points: H atom, H$_{2}^{+}$ and HHe$^{2+}$.}
\end{figure}

\begin{figure*}
\subfloat[The energy and density of HHH$^{2+}$ as the central atom is moved
at a distance R between two H atoms at 0 and 10 Angstroms. The inset
pictures show the density for several values of R. The electron is
carried across on the central proton as it moves.]{\includegraphics[angle=-90,scale=0.33]{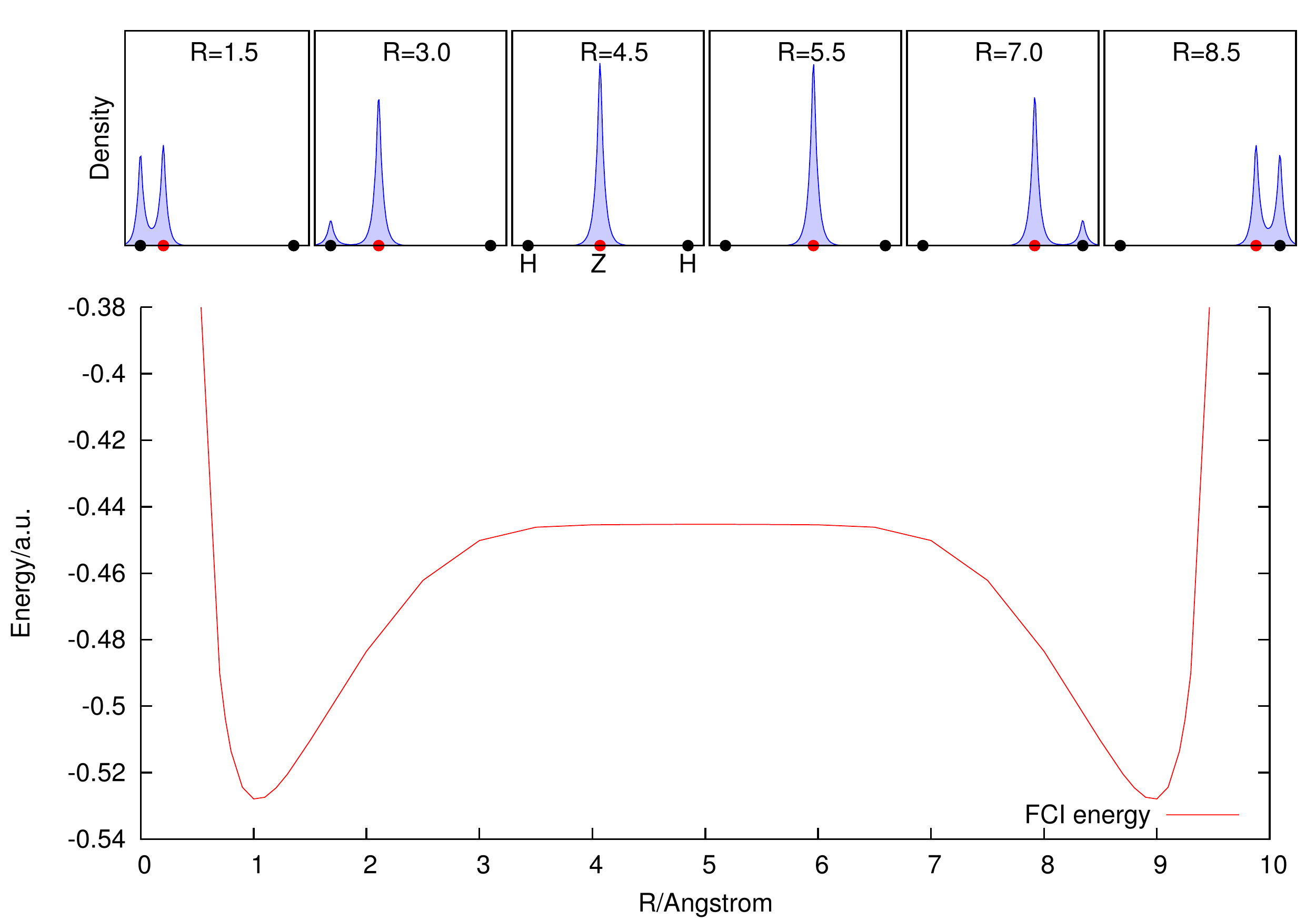}}\hfill{}\subfloat[The same as (a) but now HZH$^{1e}$ with a fractionally charged nucleus
Z=0.9 on the central atom. The inset pictures of the density show
the electron hopping between the two stationary protons as the Z proton
moves from the left to the right. ]{\includegraphics[angle=-90,scale=0.33]{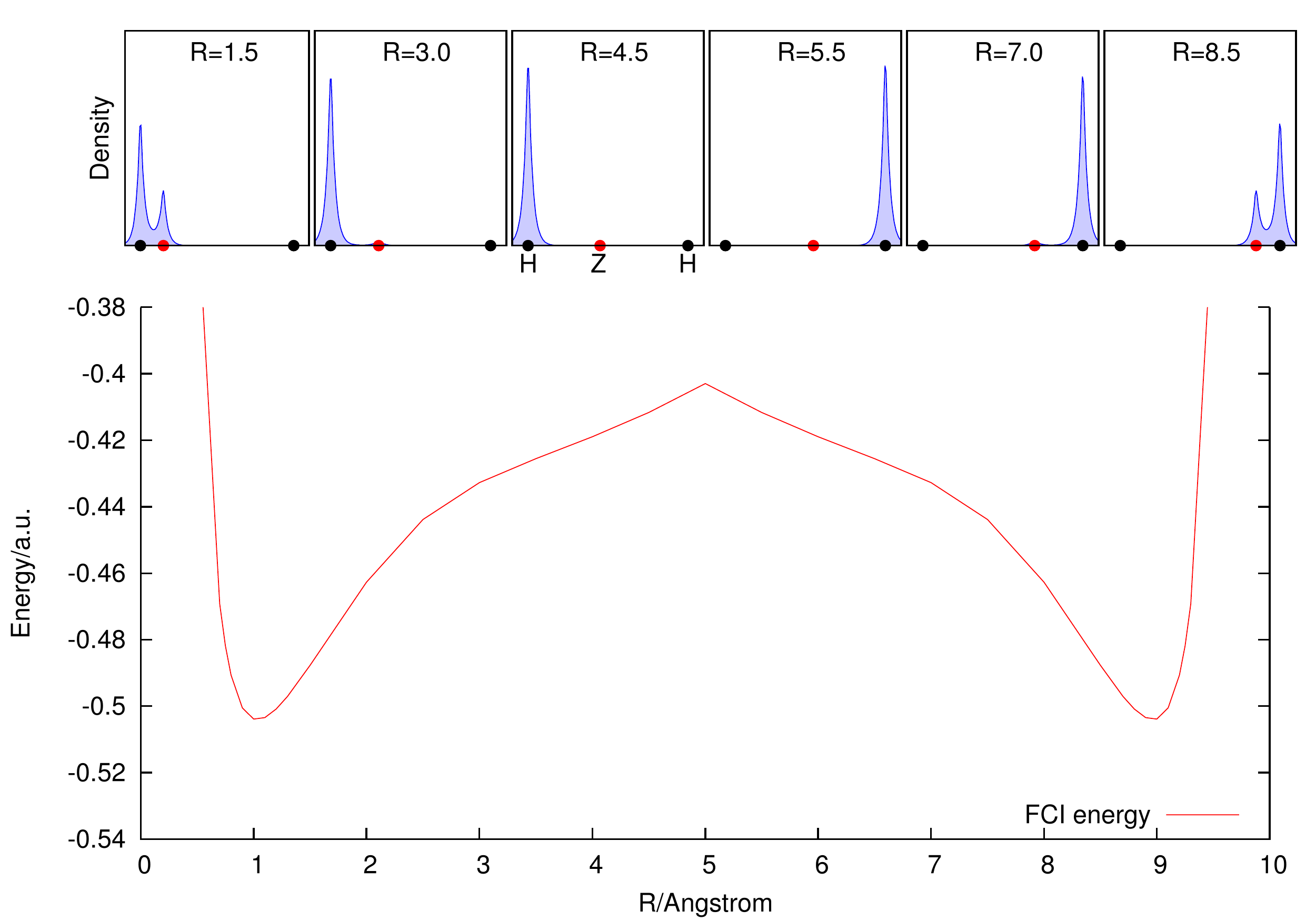}}

\caption{}
\end{figure*}

The connectedness between physical systems is investigated in this
work by taking the normal external potential for a set of nuclei $\{A\}$
at positions $\mathbf{R}_{A}$ with charge $Z_{A}$, 
\[
v(\mathbf{r})=-\sum_{A}\frac{Z_{A}}{|\mathbf{r}-\mathbf{R}_{A}|},
\]
and simply extending the realm of the charge of the nuclei from the
usual integers, $Z_{A}\in\mathbb{Z}$ to now be fractional, $Z_{A}\in\mathbb{R}$.
Fractional nuclei have been used previously in different contexts,
from alchemical changes \cite{vonLilienfeld071083,vonLilienfeld131676}
to inverse-design \cite{Keinan07176,Hu0864102} to QMMM evaluation
of pKa \cite{QMMMpKa} to density of atoms \cite{Kim13}. Here, the
idea is used to give a simple and physical controllably change to
the Hamiltonian such that, as theoreticians, a fine control over the
electronic structure problem is achieved. It should be noted that
only the external potential is changed such that electrons retain
their nature, hence all electronic structure methods should apply.
Other physically motivated ideas such as model hamiltonians \cite{CapelleReview}
like the Hubbard model or Anderson model are different in that they
also change the electron-electron interaction and, for example, conventional
DFT functionals cannot be directly applied. 

The exact behavior of electrons in some interesting but simple model
systems is now investigated using these fractional nuclear charges.
Exact calculations using full configuration interaction calculations
(FCI)\cite{FCIcode,FCIcode2} are carried out and the behavior of
the electron density $\rho(\mathbf{r})$ is examined. This reveals
basic intricacies of electronic structure and fundamental behavior
of electrons that can be seen in real space. This approach is applied
to several simple examples with one and two electrons that are able
to reveal fundamental challenges of of describing the intricate nature
of the quantum mechanical behavior of electrons. Remarkably, the visualization
of dramatic changes in the density associated to the integer nature
of electrons emerge and are possible to visualize in simple systems
only with the use of non-integer nuclear charges. 

First consider the H$_{2}^{+}$ molecule with two protons and one
electron. From a wavefunction perspective this molecule is trivial
as it only has one electron and Hartree-Fock gives the exact solution.
However, it still offers an interesting and challenging behavior that
can be illustrated by the failure of non-wavefunction methods such
as DFT, which with LDA or GGA functionals upon stretching give a massive
error in the energy of around 60 kcal/mol. This is the classic problem
of self-interaction \cite{Perdew815048,Ruzsinszky06194112} or delocalization
error \cite{Mori-Sanchez08146401} in DFT. All semi-local functionals
have a qualitative failure in the energy\emph{ }at infinity, however,
the density is not qualitatively wrong, as it is constrained by symmetry
to give half an electron on each end. In Fig. 1 we use fractional
nuclei to turn this error in the energy into an error in the density.
The charge on one of the protons (now called Z) is changed and allowed
to be non-integer. The number of electrons is always fixed, constant
at 1, hence this molecule is called HZ$^{\{1e)}$. The exact behavior
from FCI is very simple and clear at infinite separation of H and
Z. For any $Z<1$ all the electron is on the H (with corresponding
energy $-\frac{1}{2}$) and for $Z>1$ all the electron is on the
$Z$ (with corresponding energy $-\frac{Z^{2}}{2}$). For the point
at $Z=1$ ( corresponding to H$_{2}^{+})$ the electron can be found
half on the H and half on the Z. Thus, the exact behavior of the electron
density is discontinuous with respect to Z at $Z=1,$ clearly exhibiting
the integer nature of electrons. It is found that an infinitesimal
change in the Hamiltonian produces a dramatic change of the electron
density. This can be compared to the performance of a typical GGA
functional such as PBE \cite{Perdew963865}. The density for non-integer
Z reveals the delocalization error in a very visual manner. Thus,
the error for the energy of infinitely stretched H$_{2}^{+}$ is turned
into an explicit error for the density in HZ$^{\{1e\}}.$ PBE clearly
misses the discontinuity at $Z=1$ and favors a smooth charge transfer
that leads to an over delocalized electron density that is on both
H and Z. It is the first time that the delocalization error is visualized
in such a clear manner in a real space picture of a one-electron system

This simple exercise is highly illustrative of the complexity of electronic
structure that occurs even in one-electron systems. There have been
many previous papers on H$_{2}^{+},$ including many which have highlighted
qualitative failures of the energy of DFT functionals, but none of
them have focused on such an error in the density. It should be noted
that the corresponding ``chemical'' change (with no fractional nuclei),
going from H$_{2}^{+}\rightarrow$HHe$^{2+}$, does not illustrate
this failure as the error of functionals is dwarfed by the difference
in energy of the electron being on the two different atoms. From a
density functional perspective the qualitative failure is a consequence
of the delocalization error, that can be easily seen in real space.
The delocalization error in DFT implies massive failures in both the
energy and the electron density. From a chemical point of view, this
shows the particle nature of the electrons. This key aspect is missing
from currently used approximations in DFT.

Next consider the simplest possible chemical reaction with three protons
and one electron, H$_{2}^{+}+$H$^{+}\rightarrow$H$^{+}+$H$_{2}^{+}$.
For simplicity, a linear geometry is taken, with two protons fixed
10 Å apart and another proton moving between them. A 1-dimensional
coordinate, R (distance to the left proton), describes the reaction.
For R=1.0 Å the electron is near the left proton and for R=9.0 Å the
electron is near the right proton. Therefore, as the central proton
moves from the left to the right the electron w ill be transferred
as well. We also consider changing the charge on the central proton
to be fractional, giving a reaction HZ$^{\{1e\}}+$H$^{+}\rightarrow$H$^{+}+$ZH$^{\{1e\}}$.
Fig. 2(a) illustrates the hydrogen atom transfer reaction with a charge
Z=1.0. The density plots show how the electron is carried on the central
proton as it moves from left to right. However, with a charge Z=0.9
on the central proton the reaction exhibits a completely different
mechanism (Fig. 2(b)) proton transfer followed by electron transfer.
Here, as the central proton moves $1\lesssim$R$<5$ the electron
stays on the left H atom. This can easily be understood from the previous
example, HZ$^{\{1e\}}$, where the stretching leaves the electron
on the H. However, for the same reason, when R$>5$ the electron is
on the right H atom. Therefore, there is an electron hopping as R
goes through the midpoint, R=5. This is a very striking example of
a conical intersection again showing how a very small change to the
system leads to markedly different behavior of the electrons. Electronic
structure methods must be able to describe all of these mechanisms
correctly to provide a full understanding of chemical reactions and
electron transfer processes. The performance of GGA methods for these
reactions is disastrous due to delocalization error, with the electron
spread over all three centers and a corresponding unphysical drop
in energy. 

\begin{figure}
\includegraphics[bb=60bp 0bp 792bp 612bp,clip,scale=0.35]{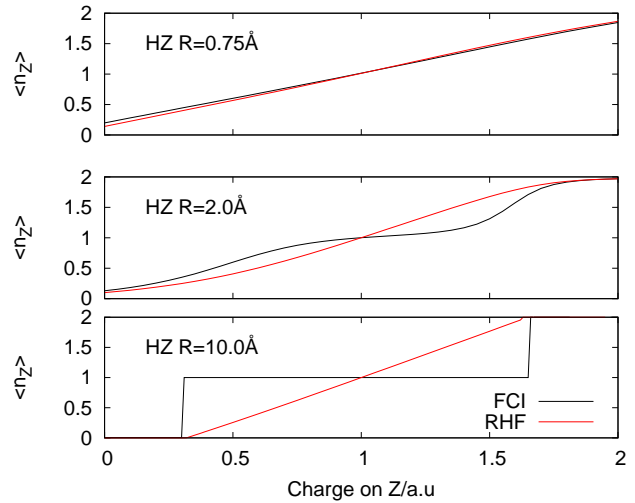}

\caption{HZ$^{\{2e\}}$ system: occupation on the Z atom ($\langle n_{Z}\rangle)$
for charge on atom Z $0\leq Z\leq2$, for three different geometries
comparing exact FCI with approximate restricted Hartree-Fock (RHF)
method.}
\end{figure}

Let us now consider closed-shell systems with two electrons. We first
study the HZ$^{\{2e\}}$ system as the nuclear charge on the Z is
varied from 0 to 2. This connects smoothly from H$^{-}$ to H$_{2}$,
and to HHe$^{+}$. At stretched geometries these three integer cases
have different occupations on the $Z$ atom, with 0, 1, and 2 electrons
respectively. The occupation of the Z atom $\langle n_{Z}\rangle$
(calculated by a simple Mulliken atomic population) is plotted in
Fig.3 as a function of the charge on the nucleus Z at three different
geometries: around equilibrium (0.75Å$\AA$), moderately stretched
(2Å$\AA$) and further stretched ($10\AA$Å). The transfer of electrons
as a function of $Z$ is different in the three distance regimes.
Restricted Hartree-Fock is able to describe the smooth transfer that
occurs at shorter distances but fails qualitatively in the stretched
limits to give any discontinuous behavior associated to electron hopping.
The same is true for any DFT method that misses the derivative discontinuity.
It does not give a step like behavior in the occupation $\langle n_{Z}\rangle$
and hence it does not correctly describe the integer nature of electrons
in such electron transfer processes. This movement of electrons encapsulated
in HZ$^{\{2e\}}$ is equivalent to that seen in the classic Anderson
model of electron transport \cite{Bergfield12066801,Stefanucci11216401},
however it offers a much simpler connection to usual chemical concepts
(real electrons and nuclei in 3-dimensional space) such that, for
example DFT approximations can easily be applied and tested.

Our model systems illustrate examples where an infinitesimal change
in the external potential leads to a drastic change in the electron
density. This can be considered from the perspective of the linear
density response function at constant number of electrons
\[
\chi(\mathbf{r},\mathbf{r}^{\prime})=\left(\frac{\delta\rho(\mathbf{r})}{\delta v(\mathbf{r}^{\prime})}\right)_{N}
\]
In most physical situations a small perturbation of the Hamiltonian
produces a small change in the electronic structure, however these
cases show interesting phenomena where a small perturbation, for example
changing the nuclear charge around $Z=1$ in the case of HZ$^{\{1e\}}$
or a movement in the geometry in HZH$^{\{1e\}}$ or a change in nuclear
charge around Z=0.3 or Z=1.7 in stretched HZ$^{\{2e\}}$ lead to dramatic
changes in the electron density. In these situations the linear response
function, $\chi$, diverges. This is a challenge for approximate theories
\cite{Hellgren12022514} as, for example, it is a priori natural to
think that a diverging $\chi$ would be very difficult for a smooth
differentiable functional of the density to reproduce. However, it
is shown here that in the one electron examples an exchange-correlation
functional such as Hartree-Fock or even a functional equal to the
negative of the Coulomb energy, $E_{xc}[\rho]=-J[\rho]$, correctly
describes a diverging linear response function. Of course any form
such as LDA or GGA do not reproduce the correct behavior or indeed
any sort of divergence of $\chi$. Another related second order response
property is the Fukui function for electron removal, $f^{-}(r)=\left(\frac{\partial\rho(\mathbf{r)}}{\partial N_{-}}\right)_{v(\mathbf{r})}=\rho^{N}(\mathbf{r})-\rho^{N-1}(\mathbf{r})$,
that for the case of HZ$^{\{2e\}}$ connects together the densities
of HZ$^{\{2e\}}$ and HZ$^{\{1e\}}$. The $f^{-}(\mathbf{r})$ around
Z=1 shows interesting and challenging behavior as even though the
density (and hence orbitals) for HZ$^{\{2e\}}$ have a smooth behavior
with no interesting features around Z=1, the Fukui function shows
up a clear discontinuous behavior due to the discontinuity in the
density in HZ$^{\{1e\}}$, which is not captured by the orbitals.
Again the usual derivative expression for the Fukui function \cite{Yang12144110}
fails completely with currently used functionals, as they are missing
the derivative discontinuity.

\begin{figure}
\includegraphics[clip,angle=-90,scale=0.35]{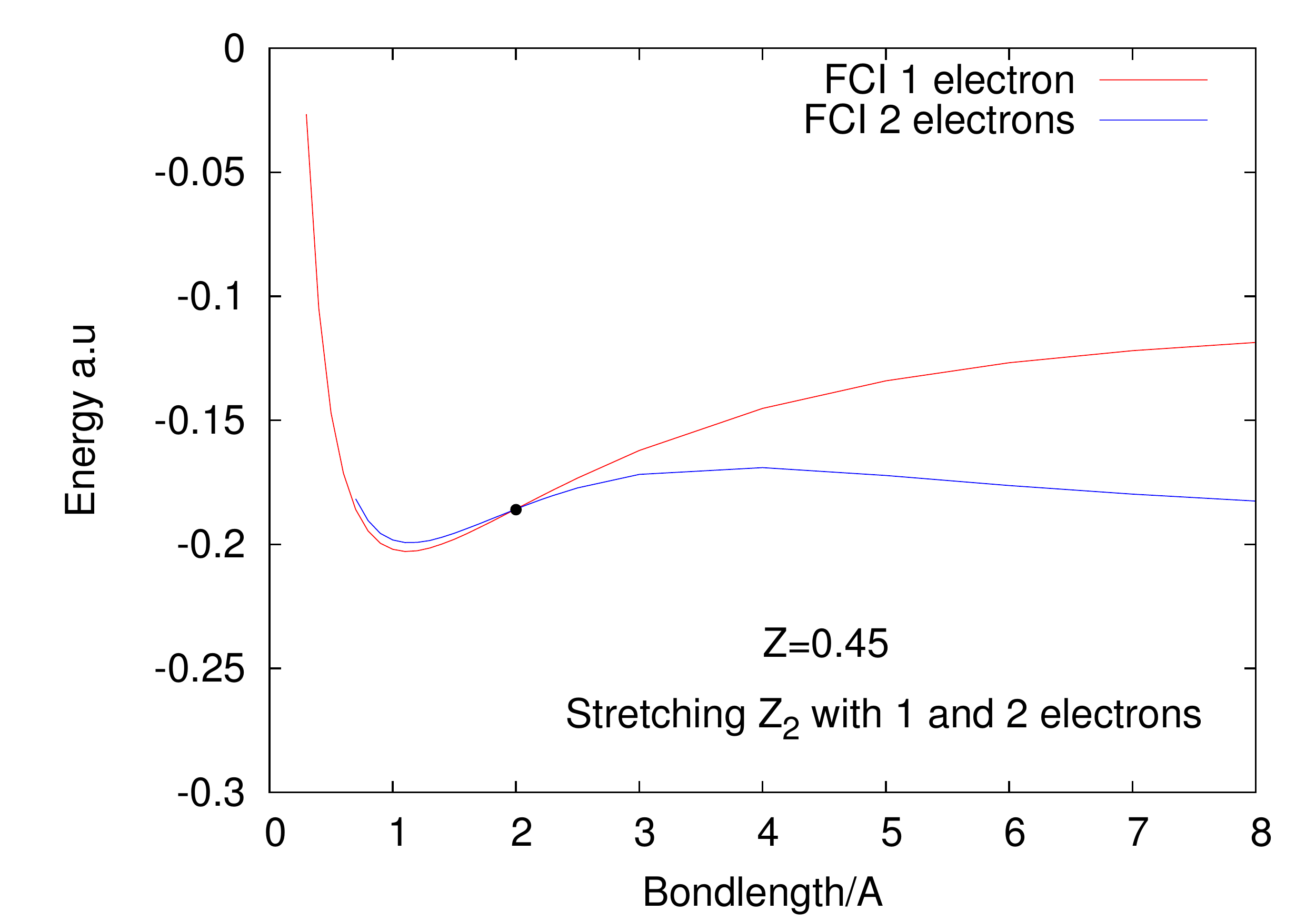}

\caption{The full CI energy with one and two electrons of the Z$_{2}$ molecule
with charge of 0.45 on both the protons calculated for different bond
lengths. There is an insulating to metallic like behavior at an internuclear
distance around 2Å, characterized by $I=A=0$ and an unbound electron.}
\end{figure}

Finally we look at the electronic structure in H$_{2}$ like molecules
and change the charge on both of the nuclei: Z$_{2}^{\{2e\}}$. Infinitely
stretched Z$_{2}^{\{2e\}}$ binds two electrons, one on each nucleus
with a total energy of $-Z^{2}E_{h}$, for all values $Z>0$. Let
us now consider the opposite extreme with zero separation, the united
atom,that is a single nucleus with charge $2Z$ that holds 2 electrons.
For example, for two Z=1 protons (i.e. H$_{2})$ the corresponding
united atom is a He atom, whereas for Z=$0.5$ protons a united atom
of H$^{-}$ is obtained. Note that the second electron in He is bound
by $-0.903E_{h}$ (the electron affinity of He$^{+})$ whereas the
second electron in H$^{-}$ is only just bound, as the electron affinity
of the H atom is now only $-0.028E_{h}$. Hence, if the nuclear charge
on the protons is further reduced to 0.45, the united atom will have
a charge of 0.9, such that now it is unable to bind two electrons.
Therefore, at some point in between $\infty$ and 0, for $Z=0.45,$
the Z$_{2}^{\{2e\}}$ system undergoes a transition from being able
to bind two electrons to only being able to bind 1 electron. This
concept is illustrated in Fig. 4 which contains the binding curves
of both Z$_{2}^{\{1e\}}$ and Z$_{2}^{\{2e\}}$ with a nuclear charge
of 0.45 from FCI calculations in a large diffuse basis set. For large
bond lengths the energy with two electrons is much lower than the
energy with one electron. However, as the bond length is decreased
the second electron becomes more weakly bound until at around 2Å the
molecule only binds one electron. This means a well characterized
transition into metallic behavior for the two electron system such
that the ionization energy is the same as the electron affinity, $I=A=0$,
and the unbound electron is released. This is one of the simplest
illustrations of an insulator to metal transition, and it cannot be
seen in H$_{2}.$ Density functional approximations such as PBE completely
fail to describe this behaviour as they have an incorrect Z$_{2}^{\{1e\}}$
curve (delocalization errror\cite{Mori-Sanchez08146401}), an incorrect
Z$_{2}^{\{2e\}}$ curve (static correlation error\cite{Cohen08121104})
and an incorrect energetic preference for Z$_{2}$ with fractional
numbers of electrons (the failure for the flat-plane \cite{Mori-Sanchez09}).

In conclusion, the foray into the theoretical of world of fractionally
charged nuclei allows us to directly visualize fundamental complexities
of electronic structure in real space. The integer nature of electrons
is critical in processes such as electron transfer or conductance
and is seen in the HZ$^{\{1e\}}$ and HZ$^{\{2e\}}$ molecules. HZ$^{\{1e\}}$
shows the particle behavior of a single electron. This is not described
by approximate methods such as DFT due to an inherent bias towards
fractional electrons leading to a delocalization error that can be
clearly seen in real space. It should be noted, however, that HZ$^{\{1e\}}$
does not capture the derivative discontinuity as it only has a single
electron and, for example, $E_{xc}[\rho]=-J[\rho]$ (a smooth differentiable
functional with no derivative discontinuity) is exact . A clear picture
of the derivative discontinuity is given by the density of stretched
HZ$^{\{2e\}}$. This shows the integer nature of two electrons that
is very challenging to describe, for example, an orbital functional
such as RHF completely fails. Electron hopping and a conical intersection
can be seen in the chemical reaction of the HZH$^{\{1e\}}$ system.
The essence of a metal-insulator transition is shown in the Z$_{2}^{\{2e\}}$
molecule for $Z=0.45$. It undergoes a geometry dependent transition
from binding two electrons to only binding one electron with a metallic
electron unbound from the nuclei, characterized by $I=A=0$. The physics
encapsulated in the behavior of electrons in all these examples is
at the heart of processes from electron transport to chemical reactions
and the insulating to metallic transition in materials. It is only
the use of fractional nuclei that reveals the full complexity of the
electronic structure offering a massive challenge for approximate
density and wave-function based methods.

We thank the Royal Society and Ramón y Cajal for funding. We dedicate
this work to the memory of Prof. N. C. Handy.

\bibliographystyle{apsrev4-1}

\end{document}